\documentclass[a4paper]{jpconf}
\usepackage{graphicx}

\def\be {\begin{equation}}
\def\ee {\end{equation}}

\begin{document}
\title{Toward a simultaneous description of $R_{AA}$ and $v_2$ for heavy quarks}

\author{S. K. Das$^{1,2}$, F. Scardina $^{1,2}$, S. Plumari$^{1,2}$ and V. Greco$^{1,2}$}

\address{$^1$ Department of Physics and Astronomy, University of Catania, Via S. Sofia 64, I-
95125 Catania, Italy}

\address{$^2$ Laboratori Nazionali del Sud, INFN-LNS, Via S. Sofia 62, I-95123 Catania, Italy}

\ead{santosh@lns.infn.it}

\begin{abstract}
The two key observables related to heavy quarks that have been measured in
RHIC and LHC energies are the nuclear suppression factor $R_{AA}$ and the elliptic flow $v_2$. 
Simultaneous description of these two observables is a top  challenge for all the existing models.  
We have highlighted how a consistent combination  of four  ingredients i.e the temperature
dependence of the energy loss, full solution of the Boltzmann collision integral for the
momentum evolution of heavy quark, hadronization by coalescence, then the 
hadronic rescattering, responsible to address a large part of such a puzzle. 
We have considered four different models to evaluate the temperature dependence of drag coefficients 
of the heavy quark in QGP. All these four different models are set to reproduce the same $R_{AA}$ 
as of the experiments. We have shown that for the same $R_{AA}$, the $v_2$ 
could be quite different depending on the interaction dynamics as well as other ingredients. 
\end{abstract}

\section{Introduction}
Theoretical calculations predict that hadronic matter at high temperatures and
densities dissolves into a deconfined state of quarks and gluons - called Quark Gluon
Plasma (QGP).The experimental efforts at Relativistic Heavy Ion Collider
(RHIC) and Large Hadron Collider (LHC) energies is aimed at to create and characterized 
the properties of QGP. The heavy hadrons (hadrons which contain at least one heavy quark, mainly c and b ) 
constitute a unique probe of the QGP properties, because they are produced in the early stage
of the collisions and they are the witness to the entire space-time evolution of the
system. 

The two key observables related to heavy quarks that have been measured in
RHIC and LHC are the nuclear suppression factor $R_{AA}$~\cite{stare,phenixelat,alice}, which is the ratio 
between the the $p_T$ spectra of heavy flavored hadrons (D and B) produced in Au+Au 
collisions with respect to those produced in p+p collisions,
and the elliptic flow $v_2$~\cite{phenixelat,alicev2}, which is a measure of the anisotropy in the angular distribution. 
Several theoretical efforts have been made to study the $R_{AA}$ and the $v_2$ measured in
experiments  within the Fokker Planck (FP) approach~\cite{rappv2,rappprl,Das,alberico,bass,Das:2015ana} and the relativistic 
Boltzmann approach~\cite{gossiauxv2,gre,fs,Song:2015sfa,fs2,Younus:2013rja}.
However all the approaches shown some difficulties to describe both the $R_{AA}$ and $v_2$ 
simultaneously. 

\section{Results}
To address the  simultaneous description of $R_{AA}$ and $v_2$, we start with the time evolution 
of $R_{AA}$ and $v_2$ to know how they develop during the expansion of the QGP.
To study the time evolution of $R_{AA}$ and $v_2$, we have solved the Fokker Planck (FP) equation 
stochastically in terms of the Langevin equation, for detail we refer to our earlier work~\cite{Das:2015ana}. 
The elastic collisions of heavy quarks with the bulk has been considered within
the framework of pQCD. The divergence associated with the t-channel diagrams due to massless 
intermediate particle exchange has been regularized introducing the Debye screening mass 
$m_D = 4\pi\alpha_s T$ with the running coupling $\alpha_s$. 
For the bulk evolution we are using a transport bulk which can reproduce some of the gross 
features of the bulk~\cite{Scardina:2012hy,Ruggieri:2013bda}. At RHIC energy, $Au+Au$ at $\sqrt{s}= 200$, we simulate the fireball 
with initial temperature  in the center is  $T_i=340$ MeV and the initial time  is $\tau_i=0.6$ fm/c. 
for the detail of the initialization, we refer to our early work. Initially the charm quark are 
distributed according to the charm production in pp collisions~\cite{initial}.

\begin{figure}[ht]
\begin{minipage}{18pc}
\includegraphics[width=14pc, clip=true]{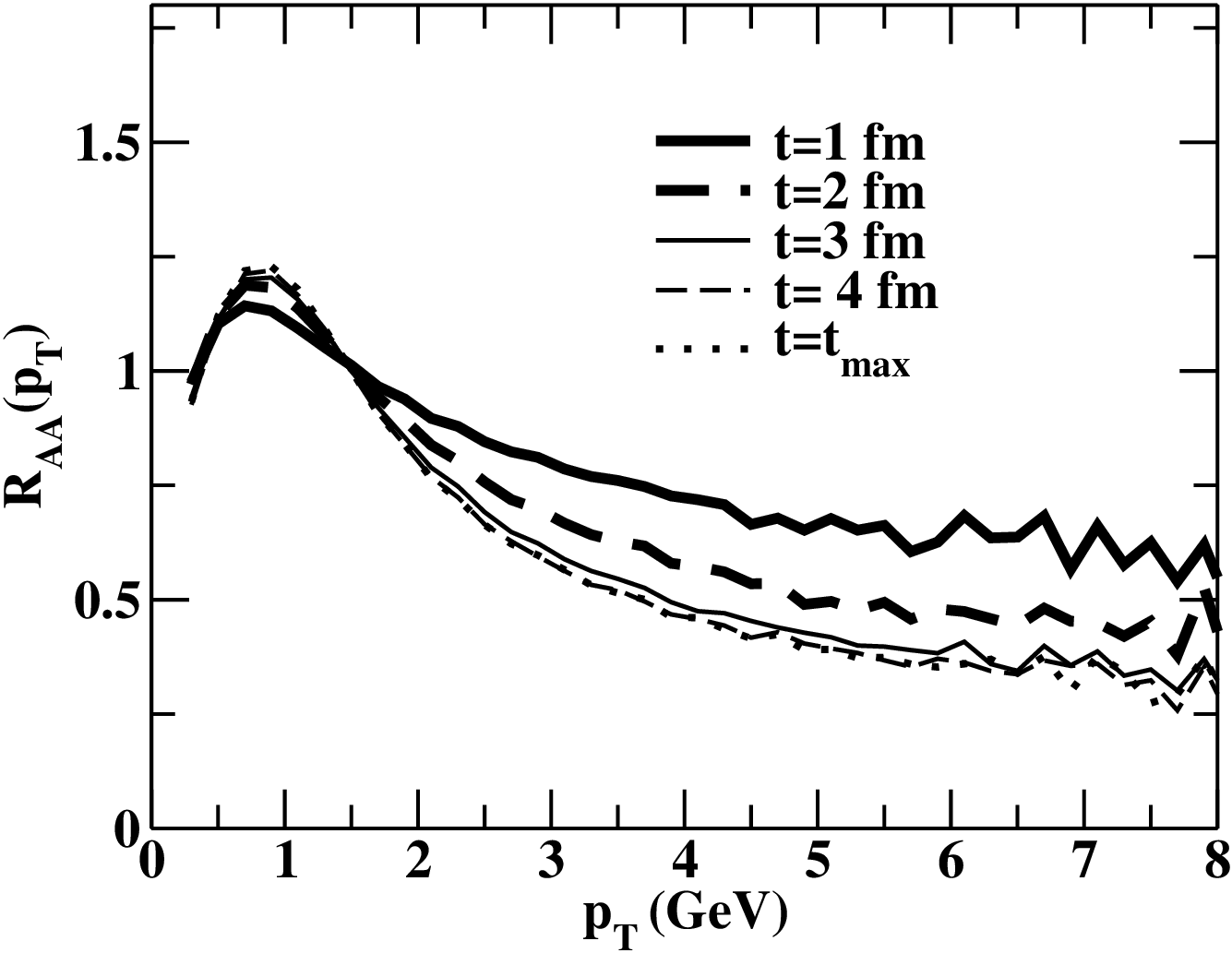}
\caption{ Time evolution of $R_{AA}$. }
\label{RAA_time}
\end{minipage}\hspace{2pc}%
\begin{minipage}{18pc}
\includegraphics[width=15pc, clip=true]{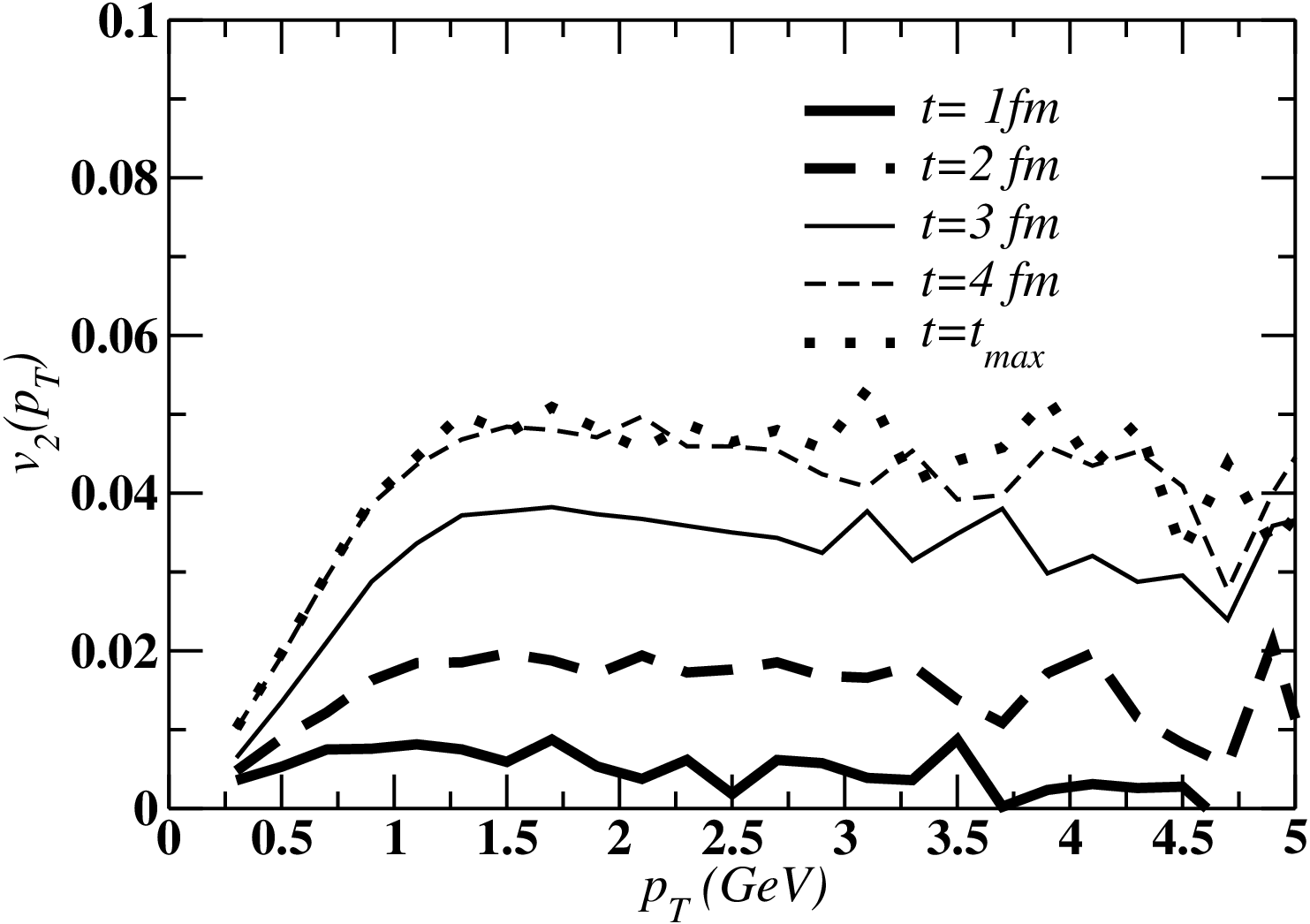}
\caption{ Time evolution of $v_2$.}
\label{v2_time}
\end{minipage} 
\end{figure}

As shown Fig~\ref{RAA_time}, the $R_{AA}$ is such which is develop at the very early stage of the 
evolution and get saturated within 3-4 fm where as the $v_2$ (in ~\ref{v2_time}) is develop at the latter 
stage of the evolution. Initially the bulk have zero $v2$. First the bulk will develop its own $v_2$ and 
then it will transfer it to the heave flavor.  
Hence the $R_{AA}$ is sensitive to the magnitude of the drag coefficient 
at the early stage of the evolution i.e at $T_i$ where as the $v_2$ is very sensitive to the 
magnitude of the drag coefficient at the latter stage of the evolution i.e at $T_c$ . 
This highlights the T-dependence of the drag coefficient may play a significance role 
for a simultaneous description of $R_{AA}$ and $v_2$  as they are sensitive to different 
range of T. To study the impact of T-dependence of drag coefficients on heavy quark observable, 
we have consider four different model to calculate the drag and diffusion coefficients 
of heavy quark in QGP.

Model-I (pQCD): The elastic collisions of heavy quarks with the bulk has been considered within
the framework of pQCD.\\
Model-II (AdS/CFT): In this case we have considered the drag force from the gauge/string duality 
i.e. AdS/CFT~\cite{Maldacena:1997re} which reads $\Gamma_{conf}= C \frac{T_{QCD}^2}{M_{HQ}}$, where $C=2.1\pm0.5$.(also see ~\cite{ali}) \\
Model-III:
In the third case, we have employed a quasi particle model (QPM) model~\cite{salvo,vc,elina} with T-dependent quasi-particle 
masses, $m_q=1/3g^2T^2$, $m_g=3/4g^2T^2$, plus 
a T-dependence background field known as bag constant, tuned to the thermodynamics of 
the lattice QCD. Such a fit lead to the coupling, $g^2(T)=\frac{48\pi^2}{[(11N_c-2N_f)ln[\lambda(\frac{T}{T_c}-\frac{T_s}{T_c})]^2}$,   
where $\lambda$=2.6 and $T/T_s$=0.57. \\
Model-IV ($\alpha_{QPM}(T),m_q=m_g=0$) 
 In this case, we are considering a model where the light quarks and gluons are massless but the 
 coupling is taken from the QPM model discussed in Model III.  \\

\begin{figure}[ht]
\begin{minipage}{12pc}
\includegraphics[width=12pc, clip=true]{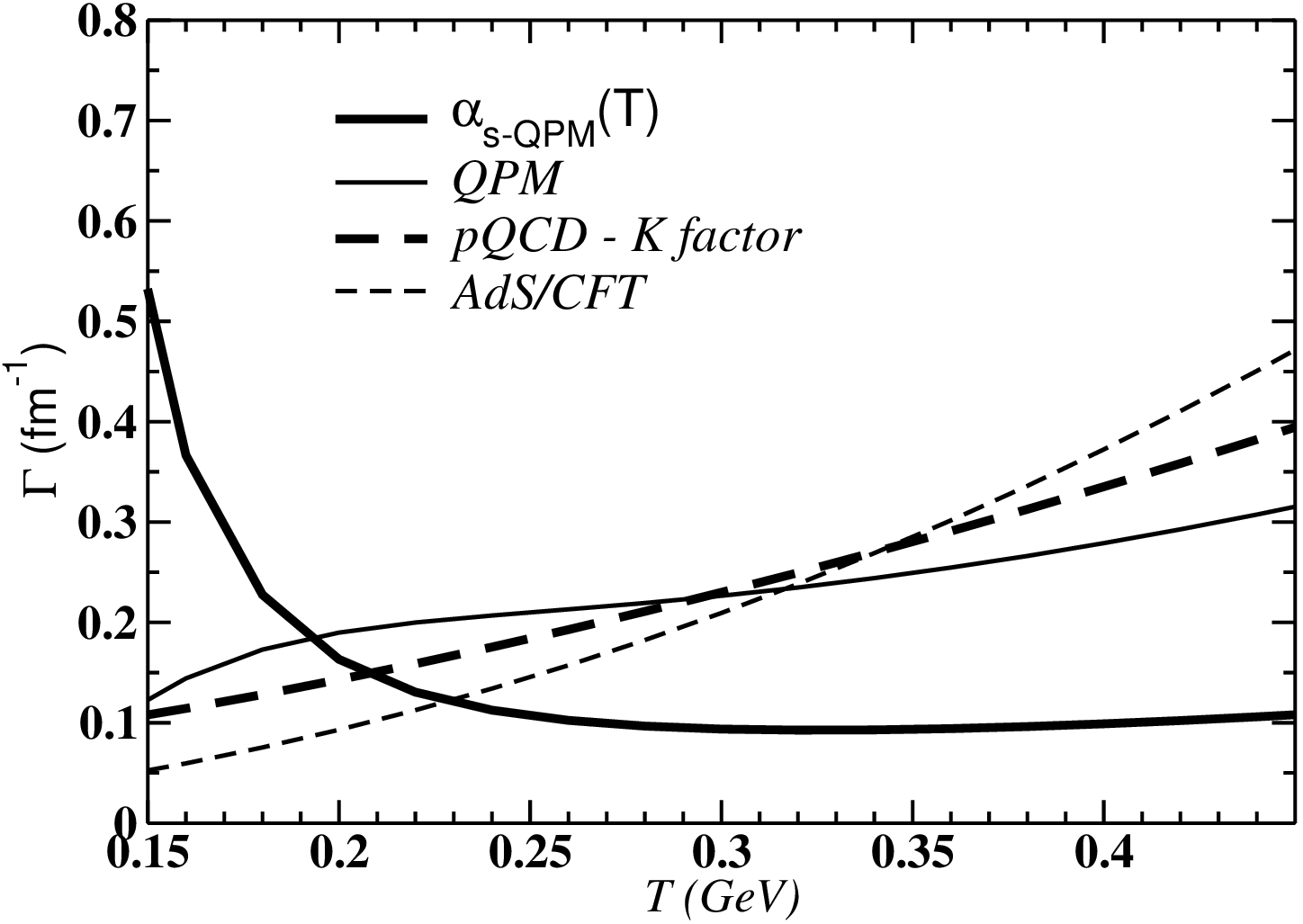}
\caption{ Drag coefficients as a function of temperature.}
\label{drag}
\end{minipage}\hspace{1pc}%
\begin{minipage}{12pc}
\includegraphics[width=12pc, clip=true]{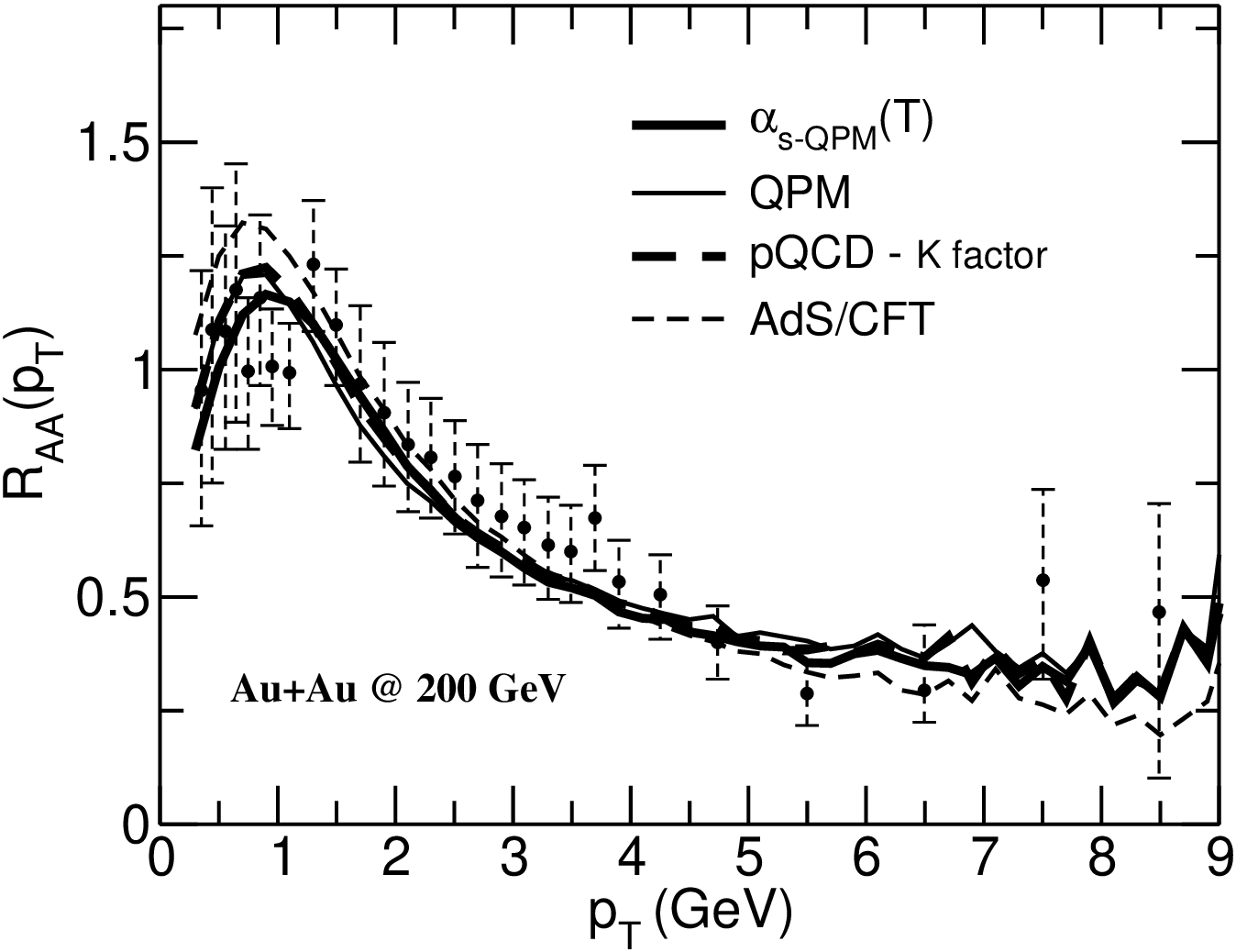}
\caption{ $R_{AA}$ as a function of $p_T$. }
\label{RAA_time1}
\end{minipage}\hspace{1pc}%
\begin{minipage}{12pc}
\includegraphics[width=12pc, clip=true]{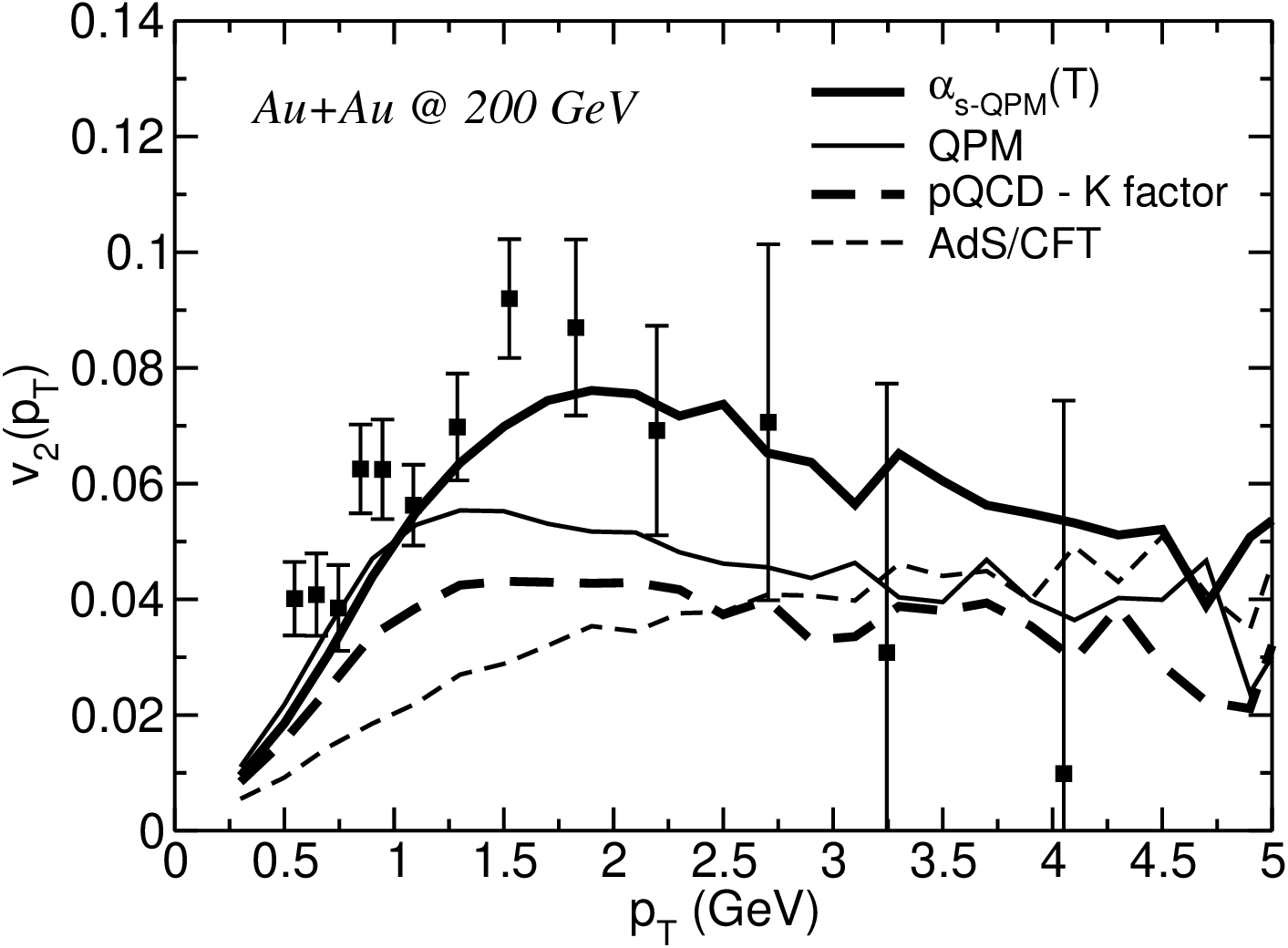}
\caption{ $v_2$ as a function of $p_T$. }
\label{v2_time1}
\end{minipage} 
\end{figure}

This fourth case has been considered to have a drag which decreasing with $T$ 
as obtained in the T-matrix approach~\cite{rappprl,riek}.The T-dependence of the drag coefficients have 
been shown in Fig~\ref{drag} obtained within the four model discussed above at $p_T=100$ MeV. 
These are the rescaled drag coefficients which 
can reproduced the same $R_{AA}$ as of the experiment at RHIC energy.

In Fig~\ref{RAA_time1} we have shown the $R_{AA}$ as a function of $p_T$ for the four
different models obtained within the Langevin dynamics at RHIC energy. The $v_2$ 
for the same $R_{AA}$ has been plotted in Fig~\ref{v2_time1} for all models as a function 
of $p_T$. We found that for the same $R_{AA}$, the $v_2$ build up can be quite different 
depending on the T-dependence of the interaction. Larger the interaction at $T_c$ larger 
is the $v_2$~\cite{Das:2015ana}. Similar effect has also been found in light quark sector~\cite{Liao:2008dk,Scardina:2010zz}. 
This suggests, the correct temperature dependence of drag coefficient has a significance role 
for a simultaneous reproduction of $R_{AA}$ and $v_2$.

\begin{figure}[ht]
\begin{minipage}{12pc}
\includegraphics[width=12pc,clip=true]{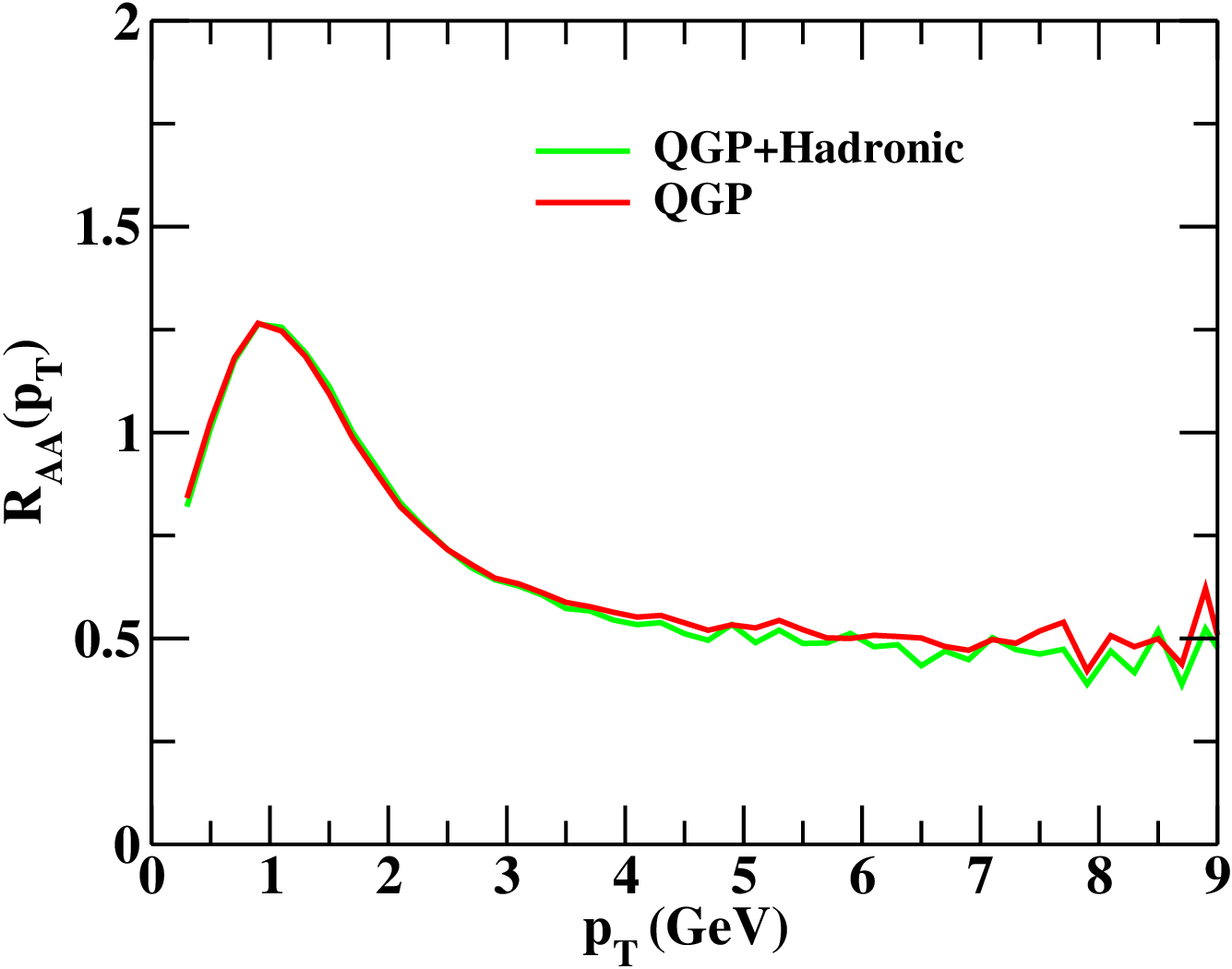}
\caption{ $R_{AA}$ as a function of $p_T$ at RHIC energy for QGP and QGP+hadronic phase.}
\label{RAA_time2}
\end{minipage}\hspace{1pc}%
\begin{minipage}{12pc}
\includegraphics[width=13pc, clip=true]{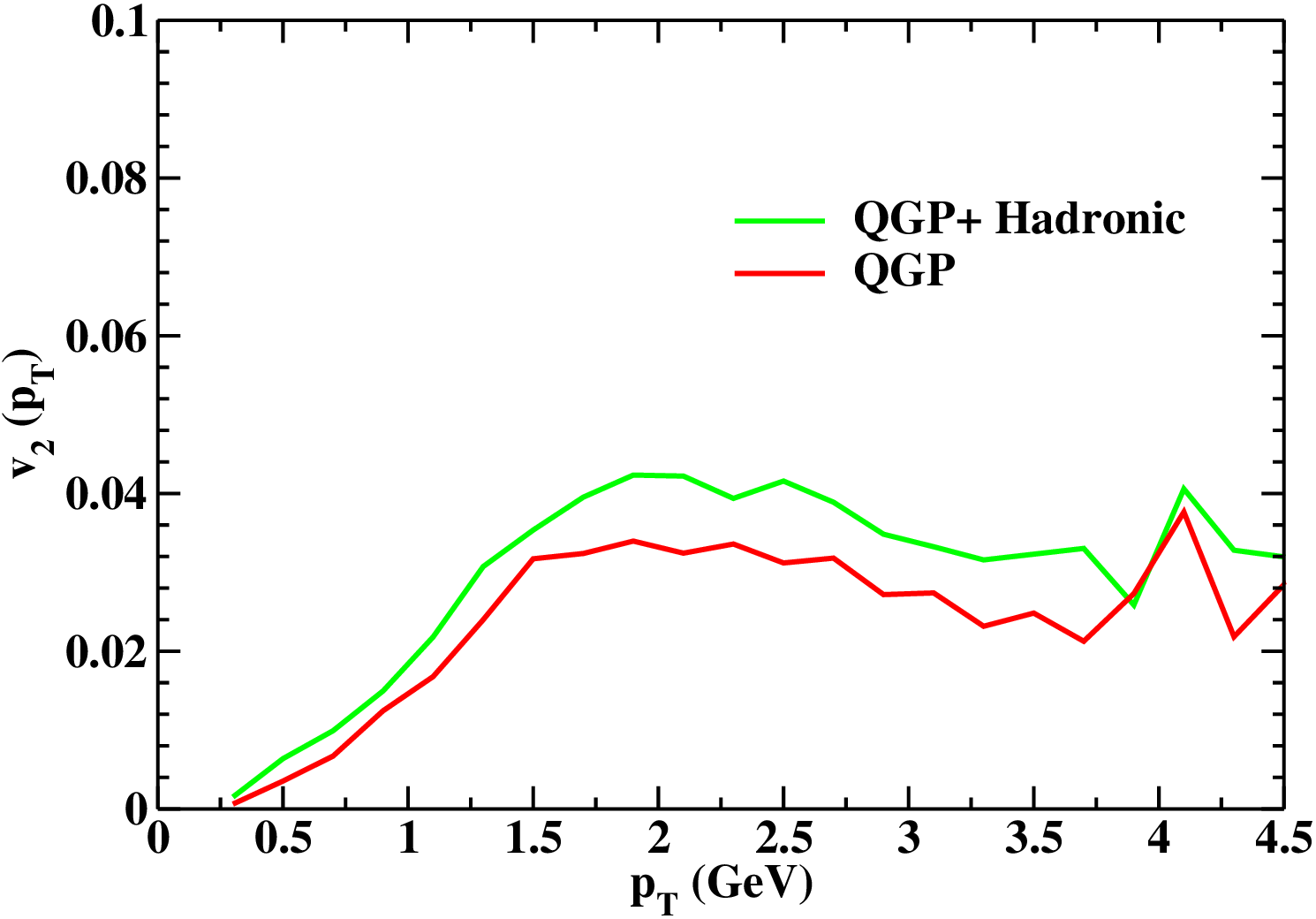}
\caption{$v_2$ as a function of $p_T$ at RHIC energy for QGP and QGP+hadronic phase. }
\label{v2_time2}
\end{minipage}\hspace{1pc}%
\begin{minipage}{12pc}
\includegraphics[width=12pc, clip=true]{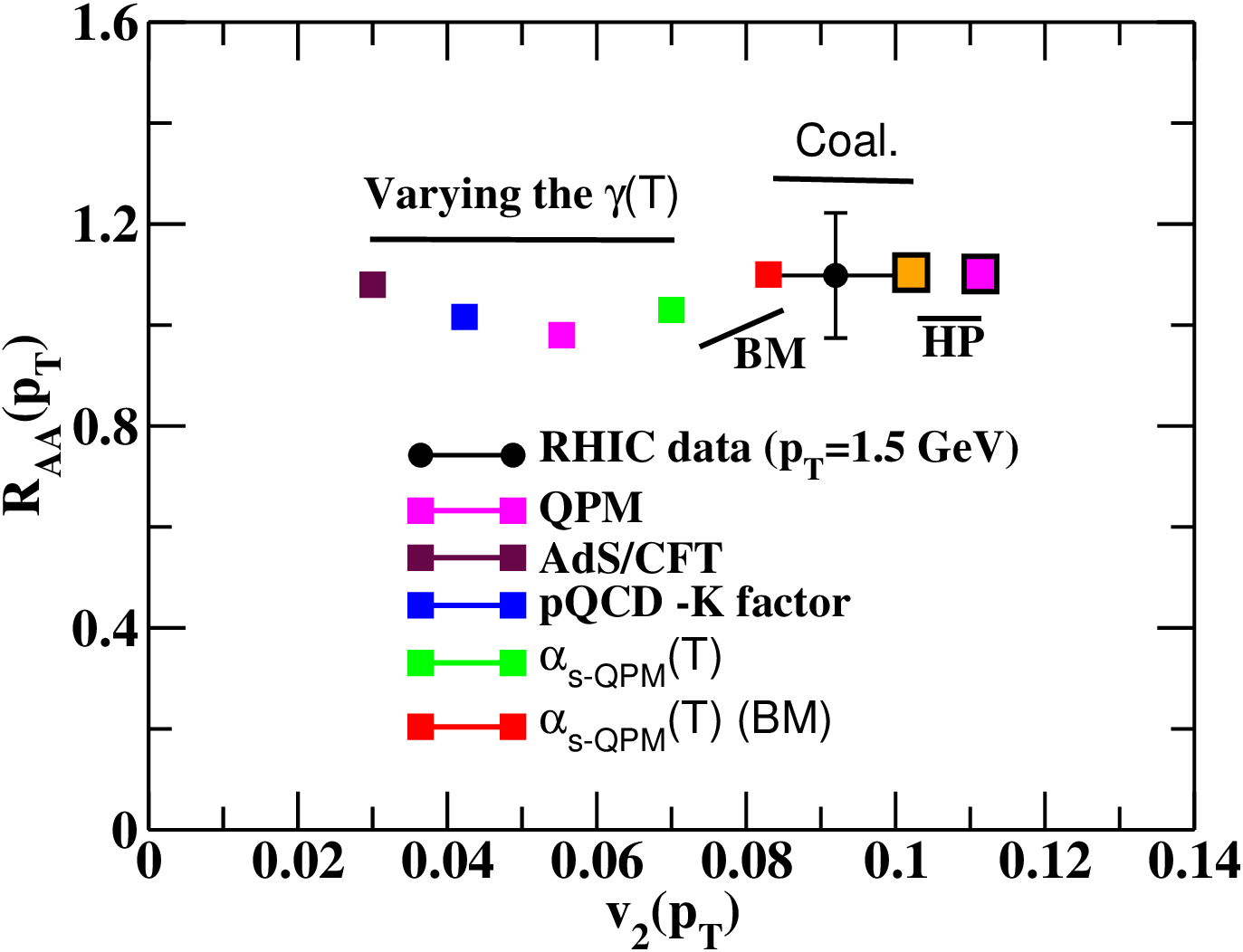}
\caption{$R_{AA}$ vs $v_2$ at $p_T=1.5$ GeV. }
\label{RAA_v2}
\end{minipage}
\end{figure}

Recently it has been shown~\cite{fs} that in case of the
full solution of the Boltzmann (BM) integral i.e. without the assumption
of small collisional exchanged momenta, leads in general to a large
$v_2$ than that of the Fokker Plank case which is a approximation of the Boltzmann 
equation under the assumption of small exchanged momenta.  Also heavy quark hadronization 
by coalescence further enhancement the $v_2$~\cite{rappprl,Greco:2003vf}. 

To study the role of the hadronic 
phase (HP) on the heavy quark observables, we study the propagation of D mesons in the 
hadronic medium consist of pions, kaons and eta. The elastic interaction between 
the D meson with the hadronic matter has been treated within the framework of chiral 
perturbation theory~\cite{hp}. In Fig~\ref{RAA_time2} we have shown the impact of the hadronic medium on $R_{AA}$ which 
is almost unnoticeable. This is because the $R_{AA}$ develop at the early stage of the 
evolution and get saturated due to radial flow within 3-4 fm. The impact of the hadronic 
medium on $v_2$ has been shown in Fig~\ref{v2_time2}. The hadronic medium further enhance 
the $v_2$ around $20\%$. In Fig~\ref{RAA_v2} we have shown how the $v_2$ build up depending on the T-dependence 
of the drag coefficients for the same $R_{AA}$. Then the $v_2$ get a boost from the 
Boltzmann equation (BM) in terms of evolution over the Fokker Planck equation. 
The $v_2$ further enhanced due to hadronization by coalescence. Finally the $v_2$ 
get the hadronic phase boost where the $R_{AA}$ remain the same.

\section{Summary and outlook}
In summary, we have shown how the $v_2$ build up for the same $R_{AA}$ depending on the 
T-dependence of the drag coefficients which is key for a simultaneous description of 
$R_{AA}$ and $v_2$. We have also highlighted how the $v_2$ get a boost from the 
Boltzmann dynamics to study the momentum evolution of heavy quark and from hadronization 
by coalescence. Then we have shown the effect of the hadronic medium on $R_{AA}$ and $v_2$.
The impact of radiative processes on heavy quark observables will also be
investigated in an upcoming study along with all these four ingredients within a single framework .
It will also be interesting to study the role of pre-equilibrium phase~\cite{Das:2015aga} on $R_{AA}$ and $v_2$ relation.

\section*{Acknowledgements} 
We acknowledge the support by the ERC StG under the QGPDyn
Grant n. 259684.

\end{document}